\documentstyle[prd,aps,doublespace,12pt]{revtex}
\begin{document}
\title{A Non-Adiabatic Controlled Not Gate for the Kane Solid State Quantum Computer}
\author{C. Wellard$^\dag$, L.C.L. Hollenberg$^{\dag\ast}$ and H.C.Pauli$^\ast$.}
\address{$^\dag$ School of Physics, University of Melbourne, 3010, AUSTRALIA.}
\address{$^\ast$ Max-Planck-Institut f\"ur Kernphysik, Heidelberg D69029, GERMANY.}
\maketitle
\begin{abstract}
\par The method of iterated resolvents is used to obtain an effective 
Hamiltonian for neighbouring qubits in the Kane solid state quantum computer. 
In contrast to the adiabatic gate processes inherent in the Kane proposal we 
show that free evolution of the qubit-qubit system, as generated by this 
effective Hamiltonian, combined with single qubit operations, is sufficient to 
produce a controlled-NOT (c-NOT) gate. Thus the usual set of universal gates 
can be obtained on the Kane quantum computer without the need for adiabatic 
switching of the controllable parameters as prescribed by Kane \cite{Kane}. Both the 
fidelity and gate time of this non-adiabatic c-NOT gate are determined 
by numerical simulation.
\end{abstract}
\newpage
\section{Introduction}
\ In the Kane proposal for a solid state quantum computer a series of spin $1\over2$ $^{31}$P nuclei in a silicon substrate are used as qubits \cite{Kane}. The interaction between these qubits is mediated by valence electrons weakly bound to the nuclei, such that at energies much lower than these electrons binding energy the system Hamiltonian is given by
\begin{eqnarray}
H &=& \mu_{B}B(\sigma_{1e}^z+\sigma_{2e}^z) - g_n\mu_n B(\sigma_{1n}^z+\sigma_{2n}^z) \nonumber\\
&+& A_1   \vec{\sigma}_{1e}. \vec{\sigma}_{1n} + A_2  \vec{\sigma}_{2e}.\vec{\sigma}_{2n} + J  \vec{\sigma}_{1e}.\vec{\sigma}_{2e}.
\label{equation:ham}
\end{eqnarray}
The interaction strengths $A_1$ , $A_2$ as well as $J$ are controllable by means of voltage biases applied to ``A'' and ``J'' gates respectively \cite{Kane,Goan}. In the case where $J=0$, the energy splitting between the electron spin states is approximately $1600$ times larger than the splitting of the nuclear states. Thus at temperatures low compared to these energy splittings the nuclear spin states, in the electron ground state, can be manipulated without significantly altering the electron states. This gives a basis in which quantum computation can be performed, the reduced Hilbert space of the electron ground state. The quantum computing basis
is then:
\begin{equation}
|\downarrow \downarrow 0 0\rangle, |\downarrow \downarrow 0 1\rangle,|\downarrow \downarrow 1 0\rangle,|\downarrow \downarrow 1 1\rangle.
\end{equation}
Here $|\downarrow\rangle$ denotes an electron in the spin down state and $|0\rangle,|1\rangle$ denote a nucleus in the spin up and down state respectively. The electron spin down state remains the ground state as long as $J < {\mu_B B \over 2}$, although the energy difference between this state and the odd superposition of electrons reduces as J increases, thus in operating our quantum computer we want to keep $J$ below this limit.
\par  A requirement of a universal quantum computer is that it be able to 
implement a universal set of gates. One such universal set comprises of the 
set of single qubit rotations and the controlled-NOT (c-NOT) gate 
\cite{Barenco1,Barenco2}. The currently proposed implementation of the c-NOT 
in the Kane quantum computer relies on an adiabatic switching on of the $A$ 
and $J$ couplings to produce a unique energy splitting between states which 
can then be swapped using a Rabi type flipping induced by a transverse 
magnetic field rotating at a frequency such that it is in resonance with the 
desired transition \cite{Kane,Goan}. This adiabatic switching process has been 
studied in detail \cite{Me} and can produce a c-NOT gate with an error of  
$\epsilon \approx 10^{-5}$ in a time $t_{\rm c-NOT} = 26 \mu s$. In this 
article we introduce a c-NOT gate that does not rely on adiabatic switching. 
We derive an effective Hamiltonian that describes the interaction of the 
qubits in the subspace of the electron ground state using the method of 
iterated resolvents. This effective Hamiltonian is then used to derive a 
c-NOT gate that relies only on the evolution of the qubits that it generates, 
in combination with the single qubit operations. We then use this fact to 
argue that the family of gates consisting of the free evolution of neighbouring 
qubits for pre determined times, is a suitable alternative to the adiabatically 
constructed c-NOT gate for use in the implementation of quantum algorithms.
\section{The Effective Hamiltonian}
\par To understand the dynamics of the qubits in the computational subspace, it is useful to calculate an effective Hamiltonian that describes the action of Eq (\ref{equation:ham}) in this reduced bases. To do this we use the method of iterated resolvents \cite{Pauli}, which has been used successfully to calculate effective Hamiltonians in reduced bases in QCD. The method of iterated resolvents involves the systematic reduction of the dimension of the system to that of the subspace of interest. In this case a reduction of the full $16 \times 16$ system Hamiltonian, $H_{16}$, to a $4 \times 4$ effective Hamiltonian, $H_4$. We begin by writing the eigenvalue problem for the complete Hamiltonian
\begin{equation}
\sum_{j=1}^{16} \langle i| H_{16} |j\rangle \langle j|\Psi\rangle = E \langle i |\Psi \rangle.
\label{equation:eval}
\end{equation}
Let us divide the rows and columns of $H_{16}$ into separate
subspaces, take the first $15$ rows to be the first subspace, call
it $P$, and the remaining subspace we call $Q$. Thus
Eq(\ref{equation:eval}) can be expressed as
\begin{equation}
\sum_{j=1}^{15} \langle i|H_{16} |j\rangle \langle j |\Psi\rangle + \langle i|H_{16} |16\rangle \langle 16 |\Psi\rangle = E \langle i | \Psi \rangle,
\end{equation}
which can be written in the block matrix form
\begin{eqnarray}
\langle P | H_{16} | P \rangle \langle P | \Psi \rangle &+& \langle P | H_{16} | Q \rangle \langle Q | \Psi \rangle = E \langle P | \Psi \rangle, \\
\label{equation:block}
\langle Q | H_{16} | P \rangle \langle P | \Psi \rangle &+& \langle Q | H_{16} | Q \rangle \langle Q | \Psi \rangle = E \langle Q | \Psi \rangle.
\end{eqnarray}
Because the eigenvalue $E$ is in general unknown, it is replaced with a free parameter $\omega$ to be determined later. Thus if the matrix $ \langle Q| \omega - H_{16} |Q\rangle$ can be inverted, the Q space wave function can be expressed in terms of the P space wave function.
\begin{equation}
\langle Q | \Psi(\omega)\rangle = G_Q (\omega) \langle Q | H_{16} |P\rangle \langle P | \Psi \rangle,
\label{equation:Qspace}
\end{equation}
where we have written the resolvent
\begin{equation}
G_Q(w) = {1 \over \langle Q|\omega - H_{16} |Q\rangle}.
\end{equation}
Substituting Eq(\ref{equation:Qspace}) into Eq(\ref{equation:block}) gives
\begin{equation}
(\langle P | H_{16}| P\rangle  + \langle P |H_{16} |Q\rangle G_Q(\omega)\langle Q| H_{16} | P \rangle) \langle P| \Psi\rangle = E(\omega)\langle P |\Psi \rangle,
\end{equation}
which defines an eigenvalue equation in the P space
\begin{equation}
\langle P |H_{15}(\omega) |P\rangle\langle P|\Psi(\omega)\rangle = E(\omega)\langle P |\Psi(\omega) \rangle,
\end{equation}
which in turn defines for an effective $15 \time 15$ Hamiltonian in the P space
\begin{equation}
H_{15}(\omega) = H_{16} + H_{16} |Q\rangle G_Q(\omega) \langle Q | H_{16}.
\end{equation}
The method of iterated resolvents calls for this procedure to be repeated until an effective $4\times4$ Hamiltonian for the computational sub-space is produced. Finally it is necessary to solve the fixed point equation $\langle4|H_4(w)|4\rangle = w$. This equation produces many solutions, the correct one to choose is the one that yields an eigenspectrum for $H_4$ that is as close as possible to the four lowest eigenvalues of the complete Hamiltonian.  This procedure was completed numerically for parameters similar to those used in the adiabatic gate $A_1 = A_2 = 1.683$ and $J = 600$ in units of $g_n \mu_n B$. This yielded an effective Hamiltonian of the form
\begin{equation}
H_{\rm eff} = \Delta (\sigma_1^x\sigma_2^x + \sigma_1^y\sigma_2^y) + \Theta \sigma_1^z\sigma_2^z + \Lambda(\sigma_1^z + \sigma_2^z) + \Gamma,
\label{equation:heff}
\end{equation}
where it is understood that the subscripts denote the first and second nucleus. The values of the parameters were $\Delta = 2.3723 \times 10^{-3} ,\Theta = -1.4645 \times 10^{-5},\Lambda = 2.6871, \Gamma = 5.3578$.
\section{Constructing a c-NOT gate}
\par The generator of the c-NOT operator, $U_{\rm c-NOT}=\exp [-i G]$, is given by
\begin{equation}
G = {\pi \over 4} ( 1- \sigma_1^z-\sigma_2^x+\sigma_1^z \sigma_2^x).
\end{equation}
All the terms on the right hand side commute, and so the c-NOT operator can be written as
\begin{equation}
{ U_{\rm c-NOT}} = {\rm exp}\bigl[-{i \pi \over 4} \bigr]{\rm exp}\bigl[{i \pi \over 4} \sigma_1^z \bigr]{\rm exp}\bigl[{i \pi \over 4} \sigma_2^x \bigr]{\rm exp}\bigl[-{i \pi \over 4} \sigma_1^z \sigma_2^x \bigr].
\label{equation:pseq}
\end{equation}
Here, reading from left to right, the first factor just represents a phase factor, this is unimportant as all that is required is a c-NOT operation up to an overall phase. The second factor is a $\sigma^z$ rotation, this can be realized by a combination of the single particle rotations common to NMR theory \cite{Gershenfeld,Ernst}.
\begin{equation}
{\rm exp} \bigl[ {i \pi \over 4} \sigma^z\bigr] = {\rm exp} \bigl[ {i \pi \over 4} \sigma^x \bigr]{\rm exp} \bigl[ -{i \pi \over 4} \sigma^y \bigr]{\rm exp} \bigl[ -{i \pi \over 4} \sigma^x \bigr].
\end{equation}
  The third factor is just another standard single particle operation. The fourth factor requires a combination of pulses to give the required evolution. Given an effective Hamiltonian of the form Eq(\ref{equation:heff}) we can use standard refocusing techniques standard to construct this evolution of the qubit-qubit system. The c-number factor in the effective Hamiltonian commutes with all other terms and simply leads to an overall phase in the evolution, it can thus be ignored in the calculations and included at the end. Our first step then is to refocus out the Zeeman evolution of the nuclei, using rf pulses targeted at both nuclei simultaneously:
\begin{eqnarray}
&{\rm exp}&[{i \pi \over 2} (\sigma_1^x + \sigma_2^x)]\ {\rm exp}[ {-i t \over 4\hbar} (H_{eff}-\Gamma)] \nonumber \times \\
&{\rm exp}&[- {i \pi \over 2} (\sigma_1^x + \sigma_2^x)]\ {\rm exp}[ {-i t \over 4 \hbar} (H_{eff}-\Gamma)] \nonumber\\
&=& {\rm exp}[ {- i t \over 2\hbar} (\Delta (\sigma_1^x\sigma_2^x + \sigma_1^y\sigma_2^y)+\Theta \sigma_1^z\sigma_2^z )].
\end{eqnarray}
The next step is to refocus out the $\sigma_1^y \sigma_2^y$ and $\sigma_1^z \sigma_2^z$ parts of the evolution:
\begin{eqnarray}
&{\rm exp}&[{i \pi \over2}\sigma_1^x]\ {\rm exp}[ {- i t \over 2\hbar} (\Delta (\sigma_1^x\sigma_2^x + \sigma_1^y\sigma_2^y)+\Theta \sigma_1^z \sigma_2^z )] \times \nonumber\\
&{\rm exp}&[-{i \pi \over 2}\sigma_1^x]\ {\rm exp}[ {- i t \over 2\hbar} (\Delta (\sigma_1^x\sigma_2^x + \sigma_1^y\sigma_2^y)+\Theta \sigma_1^z \sigma_2^z  )] \nonumber\\
&=& {\rm exp}[ {- i t \over \hbar} \Delta \sigma_1^x\sigma_2^x ].
\end{eqnarray}
Again targeting the first spin only, we can obtain
\begin{equation}
{\rm exp}[{i \pi \over 4} \sigma_1^y]\ {\rm exp}[ {- i t \over \hbar} \Delta \sigma_1^x\sigma_2^x ] \ {\rm exp}[{-i \pi \over 4} \sigma_1^y] = {\rm exp}[ {- i t \over \hbar} \Delta \sigma_1^z\sigma_2^x ];
\end{equation}
this evolution for a time $t = \pi \hbar /(4 \Delta)$ gives the required operator in Eq (\ref{equation:pseq}). Substituting these pulses into Eq(\ref{equation:pseq}) we find that the c-NOT operation can be written as,
\begin{eqnarray}
{\rm U_{\rm c-NOT}} &=& {\rm exp}\bigl[-{i \pi \over 4 }\bigl( 1-{\Gamma \over \Delta }\bigr) \bigr]{\rm exp} \bigl[ {i \pi \over 4} \sigma_1^x \bigr]{\rm exp} \bigl[ -{i \pi \over 4} \sigma_1^y \bigr]{\rm exp} \bigl[ -{i \pi \over 4} \sigma_1^x \bigr] \nonumber\\
&\times & {\rm exp}\bigl[{i \pi \over 4} \sigma_2^x \bigr]{\rm exp}\bigl[ {i \pi \over 4} \sigma_1^y\bigr] {\rm exp}\bigl[ {i \pi \over 2} \sigma_1^x\bigr]{\rm exp}\bigl[ {i \pi \over 2} (\sigma_1^x+\sigma_2^x)\bigr] \nonumber\\
&\times &F[{\pi \hbar \over 16 \Delta}] {\rm exp}\bigl[- {i \pi \over 2} (\sigma_1^x+\sigma_2^x)\bigr]F[{\pi \hbar \over 16 \Delta}]{\rm exp}\bigl[ {i \pi \over 2} \sigma_2^x\bigr]
F[{\pi \hbar \over 16 \Delta}] \nonumber\\
&\times & F[{\pi \hbar \over 16 \Delta}]{\rm exp}\bigl[ -{i \pi \over 4} \sigma_1^y\bigr],
\end{eqnarray}
where $F[t] = \exp [i \pi t H_{eff}/\hbar ]$ denotes free evolution generated by $H_{eff}$ over a time $t$. The first term is simply a phase correction, it cannot be implemented physically but tells us by what overall phase the composite operator must be corrected to produce exact c-NOT evolution.
\section{Gate Time and Fidelity}
\par It is first necessary to check that Eq(\ref{equation:heff}) does give an accurate 
description of the two qubit effective Hamiltonian. This was done by numerically solving
the Schr\"odinger equation for the time development of the entire electron-nuclear 
system using the Hamiltonian Eq(\ref{equation:ham}) over a time of several microseconds, 
and comparing the evolution of the qubits to that predicted by the effective 
Hamiltonian over the same period. It was found that final states agreed with an 
error probability of $10^{-5}$.
\par Let us now calculate the time this non-adiabatic gate takes to execute, 
using similar values for the operating parameters to those prescribed for the 
adiabatic c-NOT gate. The 
construction requires that the system evolves freely for a total time 
$t = \pi/(4 \Delta) \approx 3 \mu s$. In addition to this free evolution, 
we also have the single particle rotations, the time scale of which
is set by the time it takes to implement a $\pi$ rotation, 
$ \tau \approx 22 \mu s$. The combination of single particle operations 
necessary for this implementation takes a total of $t \approx 77 \mu s$ to 
execute, taking the total gate time to $t_{\rm c-NOT} \approx 80 \mu s$, compared 
with $26 \mu s$ for the adiabatic implementation.
\par This 
gate was simulated numerically and it was found that the operation is 
indeed that of a 
c-NOT gate, with an error probability  of $\epsilon = 4 \times 10^{-4}$. This 
is slightly outside current estimates of the error tolerance of a QC using 
error correcting codes, which vary between $\epsilon < 10^{-6}-10^{-4}$ 
\cite{Preskill,Preskill2,Aharonov}. These error tolerances, most authors 
admit, are probably pessimistic, and are based on some very general assumptions
 about the type of error process and the architecture of the computer. It is 
 possible that by tailoring an error correcting code to a specific problem, 
 the error tolerance may be more forgiving and the error probability of this 
 gate may fit well within the new bound. 
\par The problem still remains however, that 
this non-adiabatic c-NOT gate is both slower, and of lower fidelity than the 
adiabatic implementation. Both these facts can in part, be attributed 
to the large number of single qubit operations performed in the c-NOT 
implementation. Note that the period of ``free evolution'' required to 
implement the c-NOT gate is only about $3 \mu s$. During this free evolution 
is the only time that there can possibly be information flow between the two 
qubits, in contrast to the adiabatic case in which information flow occurs
over the entire gate time of $26 \mu s$. With this in mind, we can consider free evolution for this time to be 
an elementary two-qubit gate that is capable of transferring the same quantity 
of quantum information between two qubits as is a c-NOT gate, in just over $10 \%$ 
of the time. It therefore seems reasonable to speculate that it would be 
possible to recast any large scale quantum algorithm, into a series of single 
qubit rotations, and two-qubit ``free evolutions'' without the need to explicitly
construct a c-NOT gate, and that this recipe for constructing the algorithm may be
faster than one which relies on an explicit adiabatic implementation of the 
c-NOT gate, for the Kane quantum computer.
\section{Acknowledgements}
\par CJW would like to acknowledge the support of an Australian Postgraduate Award, a Melbourne University Postgraduate
Abroad Scholarship and the Max-Planck-Institut f\"ur Kernphysik. LCLH wishes to thank the Alexander Von Humboldt
foundation and the  Max-Planck-Institut f\"ur Kernphysik.

\end{document}